\documentclass[twoside]{article}
\usepackage{fleqn,espcrc2}

\newcommand{\AmS}{{\protect\the\textfont2
  A\kern-.1667em\lower.5ex\hbox{M}\kern-.125emS}}

\title{Wick quantisation of a symplectic manifold }
\author{V.A.~Dolgushev\address[TSU]{ Physics Faculty, Tomsk State University, 634050, Tomsk, Russia}
 \thanks{ On leave from Tomsk State University, Tomsk, Russia;
        ITEP, 117259 Moscow, Russia},
S.L.~Lyakhovich\addressmark[TSU] and A.A.~Sharapov\addressmark[TSU]}

\begin{document}

\begin{abstract}
{The notion of the Wick star-product is covariantly introduced for a general
symplectic manifold equipped with two transverse polarisations.
Along the lines of Fedosov method, the explicit procedure is given
to construct the Wick symbols on the manifold.
The cohomological obstruction is identified to the equivalence between the
 Wick star-product and the Fedosov one.
In particular in the K\"ahler case, the Wick star-product is shown to be
equivalent the Weyl one, iff the manifold is a Calabi-Yau one.}
\end{abstract}

\maketitle

\section{INTRODUCTION}

The progress of the quantum field theory was always related, directly
or indirectly, with in-depth study of the quantisation methods.
Nowadays we observe an explosive development in the deformation quantisation
theory. In particular, for the symplectic manifolds Fedosov
has suggested a simple method to construct a manifestly covariant
star-product \cite{Fedosov}. The Fedosov quantisation can be thought of as a
direct generalisation of the Weyl symbols known for the linear symplectic
spaces.  The Wick symbol is supposed to be
more appropriate for the field theory quantisation, being underlaid by the
notion of the creation and annihilation operators. Usually, the Fock space
construction is known for the free fields only, whereas the nonlinearity is
considered as a small perturbation to a linear background. This artificial
disintegration of the theory into the ``free" (linear) part and the
(nonlinear) ``interaction" is not always adequate to the problem, because it
may break fundamental symmetries of the model. The extended list of examples
is provided to this problem by sigma models where the nonlinear
metric can't be naturally represented as a constant ``free" part plus
``interaction".  From this standpoint, the covariant Wick symbol construction
acquires a primary importance for applying the deformation quantisation
schemes in the field theory.

In general, the Wick and the Weyl quantisations are
not equivalent to each other in the field theory even
in the linear
approximation because of possible quantum divergencies in the formal
transformation from one symbol to another. The nontrivial geometry of the
phase space may result in the cohomological obstruction to the equivalence
between different symbols (even for the system with finite number degrees of
freedom, i.e.  without any divergencies).  The explicit construction for the
Wick symbol seems to be very important to this end as well as an efficient
criterion of identifying this symbol among the other ones.

The classification of the star-products on a symplectic manifold
has been studied in several works \cite{NT}, \cite{BCG}, \cite{Deligne}\,
\cite{Xu}.
Each equivalence class of
the star-products has been shown to correspond to formal power series
(in $\hbar$) taking values in the second De Rham cohomologies.
This result seems to be most transparent  in the context of the
Fedosov method \cite{Fedosov}\,, where the space of series
in the second De Rham cohomologies
naturally appears as a moduli space of the flat Fedosov connections
of an auxiliary symbol bundle. Thus the equivalence class of the
connection defines the equivalence class of the star-product and vice versa.

In this talk we
discuss covariant construction for the Wick-type symbol
\cite{our} and identify
its equivalence class as a De Rham class of a certain 2-form.
In the K\"ahler case, the Wick and Weyl symbols turn out to be
equivalent to each other iff the manifold is the Calabi-Yau one.

In Sec 2 we briefly describe the Wick type symbol construction \cite{our}
based on the Fedosov approach \cite{Fedosov}.
In Sec 3 we study the equivalence between Wick and Weyl
symbols. Conclusion is devoted to some open questions.

\section{FEDOSOV STAR-PRODUCT OF THE WICK TYPE}

Let $(M,\{\cdot,\cdot\})$ be $2n$-dimensional Poisson manifold with Poisson
bracket
\begin{equation}\label{a1}
\{a,b\}=\omega^{ij}\partial_ia\partial_jb , \;\;\;\;\; a,b\in C^{\infty}(M).
\end{equation}
Bellow we suppose that $\omega^{ij}$ is nondegenerate, so that
inverse matrix $\omega _{ij}$ defines the symplectic 2-form
$\omega =\omega_{ij}dx^i\wedge dx^j$, which is closed in virtue of Jacobi
identity for (\ref{a1}). The commutative algebra of smooth functions
$C^{\infty}(M)$ assigned by the Poisson bracket is called
{\it Poisson algebra}
of classical observables.

The aim of deformation quantisation program is to define a new multiplication
operation $*$ (named a star-product) depending on $\hbar$,
which would be one-parametric associative deformation of the ordinary
point-wise function multiplication. More precisely,
let $C^{\infty}(M)[[\hbar]]$ be the space of formal series
$$
a=a(x,\hbar)=\sum^{\infty}_{n=0}\hbar^na_n(x),
$$
where $a_n(x)\in C^{\infty}(M)$. This space is regarded
as the space of quantum observables.
Then the star-product of two quantum observables is given by
$$
a*b=\sum^{\infty}_{n=0}{\hbar}^nD_n(a,b),
$$
where the following conditions are assumed to be satisfied:

$a)$ $D_n$ are bi-differential operators on $C^{\infty}(M)$;

$b)$ $D_0(a,b)=ab$;

$c)$ $D_1(a,b)-D_1(b,a)=-i\{a,b\}$.

The deformation quantisation, being so defined, has a natural ``gauge group"
acting on the set of all star-products. Two star-products $*$ and
$\tilde{*}$ are called equivalent if there exists an
isomorphism of algebras
\begin{equation}\label{a3}
B:(C^{\infty}(M)[[\hbar]],*)\rightarrow (C^{\infty}(M)[[\hbar]],\tilde{*})
\end{equation}
given by a formal differential operator $B =1+\hbar B_1+\hbar^2B_2+...$.

Given a Poisson manifold $(M.\{\cdot,.\cdot\})$, the problem of deformation
quantisation is to describe all the star-products satisfying axioms
$a),b),c)$ up to the equivalence (\ref{a3}).

As we have shown in \cite{our}, the most natural covariant way
to define the Wick-type  star-product is to introduce a complex-valued metric
structure $g^{ij}$ subject to certain conditions.
To represent these conditions in a compact form, let us combine the Poisson
bi-vector and the metric into the tensor field \begin{equation}
\Pi^{ij}(x)=\omega^{ij}(x)+g^{ij}(x).
\label{la}
\end{equation}
We also define a form $\Pi$ with
the lower indices as
$\Pi_{ij}=\omega_{in}\omega_{jm}\Pi^{nm}$.
The aforementioned conditions can be written as
\begin{equation}
\begin{array}{ll}
\displaystyle
i) &  {\rm rank}(\Pi_{ij})=\frac12{\rm dim}M=n  \\[6pt]
ii) & (\Pi^{in}\partial_n\Pi^{jk}-
\Pi^{jn}\partial_n\Pi^{ik})\Pi_{kn}=0
\end{array}
\label{W}
\end{equation}
These conditions guarantee the existence of two transverse
Lagrangian polarisations on $M$ associated with left and right
kernel distributions of the form $\Pi$. The integrability condition ii)
for these polarisations is also equivalent to existence of a torsion-free
Levi-Civita connection preserving $\Pi$ \cite{our}.
In what follows we will refer  to $\Pi$ as the Wick tensor.

The most notable examples of the manifolds
admitting Wick structure are provided by the K\"ahler manifolds.
In this case $\Pi^{\dagger}=-\Pi$, and the left (right) kernel
distribution of $\Pi$ generates the holomorphic (anti-holomorphic)
polarisation. The case of a real metric $g$ satisfying rels. (\ref{W})
corresponds to the so-called para-K\"ahler geometry \cite{Rocky}.

The Wick-type deformation quantisation can be defined
by replacing the axiom $c)$ with a stronger one
\begin{equation}
\begin{array}{lr}
c')& D_1(a,b)=-\frac{i}2\Pi ^{ij}(x)\partial_ia\partial_jb.
\end{array}
\label{Wick}
\end{equation}
In the paper \cite{our} the modification of the Fedosov method
was proposed to produce the $*$-product satisfying the axioms of Wick-type
quantisation.  In so doing the K\"ahler and para-K\"ahler
geometries are naturally related with the algebras of genuine Wick-
and $qp$-symbols respectively.

Let us now outline the main steps of our construction.
First, note that Wick tensor $\Pi$ on $M$ determines a constant
Wick  structure on each tangent space $T_xM$, which
can be used to quantise $T_x M$ by means of standard Wick product.
 More precisely it is expressed by

\noindent
{\bf Definition.} {\it The formal algebra $W_x$ associated to $T_xM$ is
an associative algebra with a unit over $\bf C$, whose elements are
formal power series in the deformation parameter $\hbar$ with coefficients
being formal polynomials on} $T_xM$:
$$
a(y,\hbar)=\sum_{n,m\geq 0}\hbar^{n}a_{ni_1\ldots i_m}y^{i_1}\ldots y^{i_m},
$$
{\it where y's are linear coordinates on $T_xM$. The product of elements
$a,b\in W_x$ is defined by the Wick rule}
\begin{equation}
a\circ b=exp\left(\frac{i\hbar }2\Pi^{ij}
\frac{\partial}{\partial y^i}\frac{\partial}{\partial z^j}\right) a(y,\hbar )
b(z,\hbar )|_{z=y}.
\label{wickprod}
\end{equation}
Taking a union of algebras $W_x$, $x\in M$, we obtain
a bundle $W$ of formal Wick algebras, whose sections are formal functions $$
a(x,y,\hbar)=\sum_{n,m\geq 0}\hbar^na(x)_{ni_1\ldots i_m}y^{i_1}\ldots y^{i_m},
$$
where  $a(x)_{ni_1\ldots i_m}$ are components of symmetric
covariant tensors on
$M$. The $\circ$-product  can naturally be extended to the space
$W\otimes \Lambda$ of $W$-valued differential forms by means of the usual
exterior product of the scalar forms from $\Lambda$. The general element
from $W\otimes\Lambda$ reads
\begin{equation}
\begin{array}{l}
\displaystyle
a(x,y,dx,\hbar )=\sum_{2k+p,q\ge 0}\hbar ^ka_{k\,,i_1\ldots i_pj_1.\ldots
j_q}(x) \times\\ [8pt]
\times y^{i_1}\ldots y^{i_p}dx^{j_1}\wedge \ldots \wedge dx^{j_q}.
\end{array}
\label{b1}
\end{equation}
There are two useful gradings ${\rm deg}_1$ and ${\rm deg}_2$
defined on the homogeneous elements $\hbar,y,dx$  as follows:
${\rm deg}_1(y^i)=1$, ${\rm deg}_1(\hbar)=2$, ${\rm deg}_2(dx^i)=1$,
and all other gradings  are vanishing. One may see that $W\otimes \Lambda$
is the bi-graded associative algebra with respect to the gradings
${\rm deg_1}$, ${\rm deg_2}$. The commutator of two
homogeneous forms from $W\otimes\Lambda$ is defined as
$$
[a,b]=a\circ b-(-1)^{{\rm deg}_2(a){\rm deg}_2(b)}b\circ a.
$$
Introduce the Fedosov operators
$\delta$ and $\delta^{-1}$ on $W\otimes\Lambda$, defined as follows.
For a homogeneous element $a$ with ${\rm deg}_1(a)=p$ and ${\rm deg}_2(a)=q$
we put
\begin{equation}
\delta a=dx^k\wedge \frac{\partial a}{\partial y^k}=\frac{i}{\hbar}
[\omega_{ij}y^idx^j,a],  \label{b4}
\end{equation}
\begin{equation}
\delta^{-1}a=\cases{
\displaystyle
\frac{1}{p+q}y^k i(\frac{\partial}{\partial x^k}) a,~{\rm if}~p+q \neq 0 \cr
~\cr
\displaystyle
0, ~{\rm otherwise},  }
\label{pho}
\end{equation}
Note that  both the operators are nilpotent, but only $\delta$ is the
(inner) antiderivation of $\circ$-product. The properties of these operators
are very similar to those for the usual exterior
differential and codifferential, in particular they satisfy to an
analogue of the Hodge-De Rham decomposition:
\begin{equation}
a=\sigma (a)+\delta \delta ^{-1}a+\delta ^{-1}\delta a\,,  \label{HD}
\end{equation}
where $\sigma (a)=a(x,0,0,\hbar )$ denotes the canonical projection
onto $ C^\infty (M)[[\hbar ]]$.

Now let $\nabla$  be a torsion-free connection respecting Wick tensor
$\Pi^{ij}$.
It induces the covariant derivative on
$W\otimes\Lambda$:  \begin{equation} \label{b11} \nabla a=dx^i\wedge \left(
\frac {\partial a} {\partial x^i}-y^j\Gamma _{ij}^k(x)\frac {\partial a}
{\partial y^k}\right),  \end{equation}
$\Gamma ^k_{ij}$ being the Christoffel symbols.

Following Fedosov, consider  more general
connection of the form:
\begin{equation}
D=\nabla -\delta +\frac 1{i\hbar }[r,\cdot \,],  \label{D}
\end{equation}
where $r=r_i(x,y,\hbar )dx^i\in W\otimes\Lambda ^1$\,.
Clearly, $D$ is a derivation of $\circ$-product, i.e.
\begin{equation}\label{dif}
D(a\circ b)=Da\circ b+(-1)^{deg_2(a)}a\circ Db.
\end{equation}
A simple calculations show that
\begin{equation}
D^2a=\frac 1{i\hbar }[\Omega ,a],\quad \forall a\in W\otimes \Lambda,
\label{D2} \end{equation} \begin{equation} \Omega =-\frac12\omega + R-\delta
r+\nabla r+\frac1{i\hbar }r\circ r.  \label{curve} \end{equation} Here
$R=\frac14R_{ijkl}y^iy^jdx^k\wedge dx^l$  and $R_{ijkl}=\omega_{in}R^n_{jkl}$
is the curvature tensor of $\nabla$.  A connection
of the form (\ref{D}) is called Abelian or flat if $\Omega$ does not depend
on $y's$.  In this case $D^2=0$.

Using the ``Hodge-De Rham decomposition'' (\ref{HD}), the Bianchi
identity $\nabla R=0$ and the symmetry property of the curvature tensor,
$$
\omega _{in}R^n_{jkl}=\omega
_{jn}R^n_{ikl}, \;\;\;g_{in}R^n_{jkl}=-g_{jn}R^n_{ikl},$$
 one may prove the following counterparts of  Fedosov's theorems.

\noindent
{\bf Theorem 1.}
{\it Let $\nabla$ be any torsion-free connection respecting symplectic form
$\omega=\omega_{ij}dx^i\wedge dx^j$. There is a unique
Abelian connection $D$ of the form (\ref{D}) for which}
$$
\delta^{-1}r=0,\;\;\; \Omega =\frac12\omega
$$
{\it and the expansion (\ref{b1}) for $r$ involves elements of} $deg_1\geq 3$.

Denote by $W_D$ the subspace of all parallel sections in $W$ with respect to a
flat Fedosov connection (\ref{D}). Due to rel.(\ref{dif}) the space $W_D$
is an associative subalgebra.  The next theorem establishes isomorphism
between $W_D$ and $C^{\infty}[[\hbar]]$, which induces a star-product on
$C^{\infty}[[\hbar]]$.

\noindent
{\bf Theorem 2. }{\it For any formal function
$a\in C^\infty (M)[[\hbar ]]$ there is a unique section $\widetilde{a}\in W_D$
such that $\sigma (\tilde{a})=a$. The pull-back of the $\circ$-product via
$\sigma$ induces the Wick-type star-product on} $M$:
$$
a*b=\sigma ((\sigma^{-1}a)\circ (\sigma^{-1}b)),\;\;\; \forall a,b\in
C^{\infty}(M)[[\hbar]].
$$
The elements $r\in W\otimes \Lambda^1$ and $\widetilde{a}\in W$ mentioned
in the Theorems can be efficiently constructed by iterating a pair of
coupled equations
\begin{equation}
\begin{array}{l}
\displaystyle
r=\delta ^{-1}(R+\nabla r+\frac 1{i\hbar }r\circ r),\\ [6pt]
\displaystyle
\widetilde{a} =a+\delta ^{-1}(\nabla
\widetilde{a} +\frac{1}{i\hbar} [r, \tilde{a} ])
\end{array}
\end{equation}
with respect to the first degree.

As one may see, the rank condition imposed on $\Pi^{ij}$ (\ref{W})
 is not so essential for the construction of an associative
 $*$-product obeying (\ref{Wick}). The only fact we have used
 here is the existence of a torsion-free connection preserving $\Pi $
 and the reversibility of the Poisson bi-vector $\omega^{ij}$.
 The construction would work, for example, with a
 degenerate $g$, when $g=0$ it reduces to the Fedosov quantisation.

\section{THE QUESTION OF EQUIVALENCE}

The rich geometry of symplectic manifolds equipped by the metric structure
(\ref{la}) offers at least two different schemes
for their quantisation: the Fedosov quantisation, which exploits only
antisymmetric part of the Wick form $\Pi $ (\ref{la}), and the deformation
quantisation involving
the entire form $\Pi $ to meet the condition (\ref{Wick}). The
question is whether these two quantisations are actually different
or an equivalence transform may be found
to establish a global isomorphism between both algebras of quantum
observables. Below we formulate the necessary and sufficient conditions for
such an isomorphism to exist. As in the general case, the obstruction for
equivalence of two star-products lies in the second De Rham cohomology of
symplectic manifold and we identify a certain 2-form as its representative.

In order to distinguish Wick-type star product from the Weyl one, all the
constructions related to the former product will be attributed by the
additional symbol $g$ (pointing on non-zero symmetric part $g$ in $\Pi $).
In particular, through this section the fibre-wise multiplication
(\ref{wickprod}) will be denoted by $\circ _g$, while $\circ $ will be
reserved for the Fedosov $\circ -$product \cite{Fedosov} resulting from
(\ref{wickprod}) if put $g=0$.

First we note that fibre-wise $\circ$ and $\circ _g$ products are
equivalent in the following sense:
\begin{equation}
a\circ _gb=G^{-1}(G\,a\circ G\,b),\;\;\; \forall a,b\in W \label{equq}
\end{equation}
where the formally invertible operator $G$ reads as
\begin{equation}
G=\exp\left(-\frac{i\hbar }4g^{ij}\frac \partial {\partial y^i}\frac
\partial{\partial y^j}\right). \label{G}
\end{equation}
The operator $G$ has following properties:
\begin{equation}
\nabla G=G\nabla ,\qquad \delta G=G\delta  \label{prop}
\end{equation}
Using the automorphism $G$ we can define a new Abelian connection
$\widetilde{D}=GD_gG^{-1}$,
which in virtue of rels. (\ref{prop}) can be written as
\begin{equation}
\widetilde{D}=\nabla -\delta +\frac 1{i\hbar }[\widetilde{r},\cdot ],\qquad
\widetilde{r}=Gr_g.  \label{newD}
\end{equation}
Here the brackets $[\cdot ,\cdot ]$ stand for $\circ$-commutator.
The elements $\widetilde{r}$ and $r$ satisfy the equations
\begin{eqnarray}
\widetilde{r}=\delta^{-1}(GR+\nabla \widetilde{r}-\delta
\widetilde{r}+\frac 1{i\hbar }
\widetilde{r}\circ \widetilde{r})  \label{F1} \\
r=\delta^{-1}(R+\nabla r-\delta r+\frac 1{i\hbar }r\circ r) \label{F2}
\end{eqnarray}
Thus we have two star-products $*$ and $\widetilde{*}$ corresponding to the
pair of Abelian connections $D$ and $\widetilde{D}$. Since $D\neq
\widetilde{D}$, in general, the action of the fibre-wise isomorphism $G$ establishing
the equivalence between $\circ $ and $\circ _g$-products (and hence between
the star products $\widetilde{*}$ and $*_g$) is
not automatically followed by
the equality $*=\widetilde{*}$ .
Indeed, evaluating lowest orders in $\hbar $
we get:
\begin{equation}
\begin{array}{l}
\displaystyle
a*b =ab+\frac{i\hbar }2\omega ^{ij}\nabla _ia\nabla _jb-\\ [6pt]
\displaystyle
-\frac{\hbar ^2} 2\omega ^{ik}\omega ^{jl}
\nabla _i\nabla _ja\nabla _k\nabla _lb+{\cal O}(\hbar ^3),\\ [6pt]
\displaystyle
a\widetilde{*}b = ab+\frac{i\hbar }2\widetilde{\omega }^{ij}\nabla
_ia\nabla _jb- \\ [6pt]
\displaystyle
\frac{\hbar ^2}2\omega ^{ik}\omega ^{jl}\nabla _i\nabla
_ja\nabla _k\nabla _lb+{\cal O}(\hbar ^3),
\end{array}
\label{sts}
\end{equation}
where
\begin{equation}
\begin{array}{l}
\displaystyle
\widetilde{\omega }^{ij}=\omega ^{ij}+\hbar \omega _1^{ij},  \;\;\;
\omega _1^{ij}=\omega ^{ik}\Omega _{kl}\omega ^{lj},\\ [6pt]
\Omega =\frac
i\hbar (GR-R)=\frac 18R_{ijkl}g^{ij}dx^k\wedge dx^l
\end{array}
\label{def}
\end{equation}
The 2-form $\Omega $ is closed in virtue of the Bianchi identity for the
curvature tensor. In fact, rels. (\ref{sts}), (\ref{def}) show that the
second star product $\widetilde{*}$ is a so-called {\it 1-differentiable
deformation }\cite{Stern} of $*$ . This deformation is
known to be trivial iff the 2-form $\Omega $ is exact \cite{Stern}.
Now supposing $\Omega =d\psi $ let us try to establish
an equivalence between $*$ and $\widetilde{*}$ by means of a fibre-wise
conjugation automorphism

\begin{equation}
a\rightarrow U\circ a\circ U^{-1},  \label{iU}
\end{equation}
where $U$ is an invertible element of $W$. The element $U$ is so chosen
that transformation (\ref{iU}) turns $D$ to $\widetilde{D}$.
This  is equivalent to
\begin{equation} D(U\circ a\circ U^{-1})=U\circ (\widetilde{D}a)\circ
U^{-1},\quad \forall a\in W \end{equation} The last condition means that
\begin{equation}
U^{-1}\circ DU=\frac 1{i\hbar }\Delta r+\frac 1{i\hbar }\psi ,  \label{urav}
\end{equation}
where $\Delta r=\widetilde{r}-r$ and $\psi $ is a globally defined 1-form on
$M$.  The compatibility condition for equation (\ref{urav}) resulting from
the identity $D^2=0$ requires that
\begin{equation} D\Delta r+
\frac 1{i\hbar}\Delta r\circ \Delta r+d\psi =0
\label{sovm}
\end{equation}
The analogous
relation is obtained if we subtract (\ref{F1}) from (\ref {F2})
\begin{equation}
D\Delta r+\frac 1{i\hbar }\Delta r\circ \Delta r+\Omega =0  \label{compat}
\end{equation}
Comparing (\ref{sovm}) with (\ref{compat}) we conclude that the
compatibility condition holds provided $\Omega $ is exact. Now rewrite (\ref
{urav}) in the form
\begin{equation}
\delta U=\nabla U+\frac 1{i\hbar }[r,U]-\frac 1{i\hbar }U\circ (\Delta
r+\psi )  \label{urav2}
\end{equation}
and apply the operator $\delta ^{-1}$ to both sides of the equation. Using
the Hodge-De Rham decomposition (\ref{HD}) and taking $\sigma (U)=1,$ we
get
\begin{equation}
U=1+\delta ^{-1}(\nabla U+\frac 1{i\hbar }[r,U]-\frac 1{i\hbar }U\circ
(\Delta r+\psi ))  \label{proce}
\end{equation}
The iterations of the last
equation yield a unique solution for (\ref{urav2}) provided the
compatibility condition (\ref{sovm}) is fulfilled$.$ Starting from 1, this
solution defines an invertible element of $W$. Then the
equivalence transform $B:(C^\infty (M)[[\hbar]],*)\rightarrow
(C^\infty (M)[[\hbar]],*_g)$ we
are looking for is defined as the sequence of maps
\begin{equation}
Ba(x)=(U\circ G(\sigma _g^{-1}(a))\circ U^{-1})|_{y=0},  \label{on}
\end{equation}
so that
\begin{equation}
a*_gb=B^{-1}((Ba)*(Bb))
\end{equation}
Thus we have prove the following

\noindent
{\bf Theorem 3.}{\it The obstruction to equivalence between Weyl and Wick
type deformation quantisations lies in the second De Rham cohomology }$
H^2(M) ${\it . The quantisations are equivalent iff the 2-form }$
R_{ijkl}g^{ij}dx^k\wedge dx^l${\it \ is exact}.

For the anti-Hermitian matrix $\Pi $ the 2-form $\Omega $ is
nothing but the Ricci form of the K\"ahler manifold. In this
case the cohomology class of $\Omega $, being proportional to the first
Chern class $c_1(M)$, is known to depend only on the complex structure of
the manifold \cite{Chern}. Since, for example, $c_1({\bf C}P^n)\neq 0$
and for any K\"ahler manifold $M$ a topological equivalence $M\sim
{\bf C}P^n$ implies a bi-holomorphic one \cite{Ko-Hi}, the Weyl and Wick
quantisations on ${\bf C}P^n$ are not equivalent for any $\Pi $.
More generally, the K\"ahler manifolds with the vanishing first Chern
class are known as the Calabi-Yau ones. These manifolds have been
intensively studied by physicists
in the context of string compactification problem during the last decades.
The above Theorem allows one to characterize these manifolds
as those K\"ahler manifolds on which Weyl and Wick quantisations
are equivalent to each other. This observation might be
crucial for the consistent string compactification on
the Calabi-Yau in the presence of non-constant background
B-field \cite{CDS}, \cite{SW}.

\section{CONCLUDING REMARKS}
In this talk we have explained a method to covariantly construct Wick-type
symbol for the system whose phase space is equipped with both the symplectic
structure and the metric one, interrelated by the conditions
(\ref{la}), (\ref{W}).
We have also described the relationship between Weyl
and Wick symbols and the cohomological obstructions to their equivalence.
Let us briefly discuss how this construction could be used in quantising
field theories. First, the quantisation scheme should be extended to the
constrained Hamiltonian systems, as the phase space of strings and gauge
fields is subject to constraints. A step has been done in this direction in
the paper \cite{diracbra}, where the Weyl deformation quantisation
is worked out for the Dirac brackets. Now this should be combined with the
Wick structure of the constrained phase space.
Second, it should be understood what could  be taken as a natural symmetric
{\it phase space} tensor $g_{ij}$ (\ref{la}),(\ref{W}) for the
constrained systems like strings in curved space-time.
A possible way of finding this structure is to induce it from
an enveloping linear phase space (where one usually has an obvious
Wick structure) to the nonlinear constrained surface.
It is not quite obvious, however, that the constrained surface would be
equipped in this way by an integrable Wick structure.
If the integrability condition (ii) (\ref{W}) was not
automatically satisfied for the induced Wick structure, one may hope to
overcome this obstacle by introducing an appropriate
torsion.
This can probably be done in the way of the recent paper
\cite{KarS}  where our construction of the Wick-type star-product \cite{our}
is generalised to the case of the almost K\"ahler manifold.

{\bf Acknowledgments}.
We are grateful to I.A.~Batalin, A.~Borowiec,  M.~Grigoriev, A.~Karabegov,
R.~Marnelius for valuable discussions related to the topic of this talk.
The work is partially supported by the RFBR grant No 00-02-17-956.
VAD is supported by grant for Support of Scientific Schools N
00-15-96557, SLL and AAS acknowledge the support from the grant
E-00-33-184 from Russian Ministry of Education.

\end{document}